\documentclass[a4paper,12pt]{article}
\usepackage{mathptm}
\usepackage{setspace}
\usepackage{color}
\usepackage{psfig}
\usepackage{pdfcolmk}   
\usepackage{multicol}

\newcommand{\degC}{{\,}^\circ C}
\newcommand{\be}{\begin{equation}}
\newcommand{\ee}{\end{equation}}
\newcommand{\bea}{\begin{eqnarray}}
\newcommand{\eea}{\end{eqnarray}}

\def\url#1{\textcolor{blue}{\underline{#1}}}	

\def\textformat{ 0 }	

\def\preprint#1{	
 \ifnum \textformat = 0
 \else
   #1
 \fi
}

\begin{document}

\renewcommand{\baselinestretch}{1.1} \large\small		
\preprint{
\renewcommand{\baselinestretch}{1.5} \large\normalsize	
}



\title{Extracting synaptic conductances \\ from single membrane
potential traces}

\author{\ \\ Martin Pospischil, Zuzanna Piwkowska, Thierry Bal and
Alain Destexhe* \\ \ \\ \ \\ Integrative and Computational
Neuroscience Unit (UNIC), \\ UPR-2191, CNRS, Gif-sur-Yvette, France
\\ \ \\ \ \\ $*$ Corresponding author at the address above \\ Email:
Destexhe@iaf.cnrs-gif.fr \\ Tel: +33 1 69 82 34 35 \\ \ \\ \ \\
{\it Neuroscience}, in press. \\ \preprint{Copy also available
at the arXiv preprint server: \\ \url{http://arxiv.org/abs/0807.3238}
\\ } }

\maketitle



\preprint{
\clearpage
\section*{Abbreviations}
SD = standard deviation \\
V$_m$ = membrane potential
}



\preprint{\clearpage}
\section*{Abstract}

In awake animals, the activity of the cerebral cortex is highly
complex, with neurons firing irregularly with apparent Poisson
statistics.  One way to characterize this complexity is to take
advantage of the high interconnectivity of cerebral cortex and use
intracellular recordings of cortical neurons, which contain
information about the activity of thousands of other cortical
neurons.  Identifying the membrane potential (V$_m$) to a stochastic
process enables the extraction of important statistical signatures of
this complex synaptic activity.  Typically, one estimates the total
synaptic conductances (excitatory and inhibitory) but this type of
estimation requires at least two V$_m$ levels and therefore cannot be
applied to single V$_m$ traces.  We propose here a method to extract
excitatory and inhibitory conductances (mean and variance) from
single V$_m$ traces.  This ``VmT method'' estimates conductance
parameters using maximum likelihood criteria, under the assumption
that synaptic conductances are described by Gaussian stochastic
processes and are integrated by a passive leaky membrane.  The method
is illustrated using models and is tested on guinea-pig visual cortex
neurons {\it in vitro} using dynamic-clamp experiments.  The VmT
method holds promises for extracting conductances from single-trial
measurements, which has a high potential for {\it in vivo}
applications.

\



\preprint{
\section*{Keywords} 
{\it Computational models; Cerebral cortex; Synaptic noise; 
Conductance estimation; Maximum likelihood; Dynamic-clamp}
}



\clearpage

In awake animals, neurons in different cortical structures display
highly irregular spontaneous firing (Evarts, 1964; Steriade and
McCarley, 1990).  This level of firing must be considered together
with the dense connectivity of cerebral cortex.  Each pyramidal
neuron receives between 5000 and 60000 synaptic contacts and a large
part of this connectivity originates from the cortex itself (DeFelipe
and Farinas, 1992; Braitenberg and Sch\"uz, 1998).  As a consequence,
many synaptic inputs are simultaneously active onto cortical neurons
in intact networks.  Indeed, intracellular recordings in awake
animals reveal that cortical neurons are subjected to an intense
synaptic bombardment and, as a result, are depolarized and have a low
input resistance (Matsumara et al., 1988; Baranyi et al., 1993;
Steriade et al., 2001) compared to brain slices kept {\it in vitro}. 
This activity is also responsible for a considerable amount of
subthreshold fluctuations, called ``synaptic noise''.  This noise
level and its associated ``high-conductance state'' greatly affect
the integrative properties of neurons (reviewed in Destexhe et al.,
2003; Destexhe, 2007; Longtin, 2008).

To characterize this activity, one must use appropriate methods to
measure not only the mean conductance level of excitatory and
inhibitory inputs, but also their level of fluctuations which is
quantified by the conductance variance.  Many methods exist to
extract mean conductances (see review by Monier et al., 2008), but
few methods have been proposed to extract the variances.  One class
of methods consists of identifying the membrane potential (V$_m$) to
a multidimensional stochastic process (Rudolph and Destexhe, 2003c,
2005).  The so-called ``VmD method'' (Rudolph et al., 2004) enables
the extraction of mean and variance of excitatory and inhibitory
conductances.  This method, however, requires to use at least two
different levels of V$_m$ activity, which prevents its application to
single-trial measurements.

More recently, a related method was proposed to extract
spike-triggered average conductances from the V$_m$ (Pospischil et
al., 2007).  In this case, one needs to extract synaptic conductances
time courses from a single trace (the spike-triggered V$_m$), which
was made possible using a maximum likelihood method.  

In the present paper, we attempt to merge these concepts and provide
a method to estimate the excitatory and inhibitory conductances and
their variances.  This ``VmT'' method provides similar estimates as
the VmD method, but it does so by using a maximum likelihood
estimation, and thus can be applied to single V$_m$ traces.



\section*{Experimental procedures}

\subsection*{Models}

All methods and analyses shown in this paper are based on a
stochastic model of synaptic background activity, which was modeled
by fluctuating conductances (Destexhe et al., 2001).  In this model,
the synaptic conductances are stochastic processes, which in turn
influence V$_m$ dynamics.  According to this ``point-conductance''
model,  the membrane potential dynamics is described by the following
set of equations: 
\bea
 C \ \frac{dV}{dt}  & = & - G_L \, (V-E_L)
    - g_e \, (V-E_e) - g_i \, (V-E_i) + \, I_{ext} ~ , \label{PCmodel} \\
 \frac{dg_e(t)}{dt} & = & 
 - \frac{1}{\tau_e} \, [ g_e(t) - g_{e0} ]
 + \sqrt{\frac{2 \sigma^2_e}{\tau_e}} \ \xi_e(t) ~ ,    \label{flucte} \\
 \frac{dg_i(t)}{dt} & = & 
 - \frac{1}{\tau_i} \, [ g_i(t) - g_{i0} ]
 + \sqrt{\frac{2 \sigma^2_i}{\tau_i}} \ \xi_i(t) ~ , \label{flucti}
\eea
where $C$ denotes the membrane capacitance, $I_{ext}$ a stimulation
current, $G_L$ the leak conductance and $E_L$ the leak reversal
potential.  $g_e(t)$ and $g_i(t)$ are stochastic excitatory and
inhibitory conductances with respective reversal potentials $E_e$ and
$E_i$.  The excitatory synaptic conductance is described by
Ornstein-Uhlenbeck stochastic processes (Eq.~\ref{flucte}), where
$g_{e0}$ and $\sigma^2_e$ are, respectively, the mean value and
variance of the excitatory conductance, $\tau_e$ is the excitatory
time constant, and $\xi_e(t)$ is a Gaussian white noise source with
zero mean and unit standard deviation. The inhibitory conductance
$g_i(t)$ is described by an equivalent equation (Eq.~\ref{flucti})
with parameters $g_{i0}$, $\sigma^2_i$, $\tau_i$ and noise source
$\xi_i(t)$.  For details about Ornstein-Uhlenbeck stochastic
processes, see the review by Gillespie (1996).

This model was simulated together with a leaky integrator mechanism
with no spiking mechanism because the method only applies to
subthreshold activity.  All numerical simulations were performed in
the NEURON simulation environment (Hines and Carnevale, 1997) and
were run on PC-based workstations under the LINUX operating system.

\subsection*{Experiments}

{\it In vitro} experiments were performed on 0.4~mm thick coronal or
sagittal slices from the lateral portions of guinea-pig occipital
cortex.  Guinea-pigs, 4-12 weeks old (CPA, Olivet, France), were
anesthetized with sodium pentobarbital (30~mg/kg).  The slices were
maintained in an interface style recording chamber at 33-35$\degC$. 
Slices were prepared on a DSK microslicer (Ted Pella Inc., Redding,
CA) in a slice solution in which the NaCl was replaced with sucrose
while maintaining an osmolarity of 307~mOsm.  During recording, the
slices were incubated in slice solution containing (in mM): NaCl,
124; KCl, 2.5; MgSO$_4$, 1.2; NaHPO$_4$, 1.25; CaCl$_2$, 2;
NaHCO$_3$, 26; dextrose, 10, and aerated with 95\% O$_2$, 5\% CO$_2$
to a final pH of 7.4.  Intracellular recordings following two hours
of recovery were performed in deep layers (layer IV, V and VI) in
electrophysiologically identified regular spiking and intrinsically
bursting cells.  Electrodes for intracellular recordings were made on
a Sutter Instruments P-87 micropipette puller from medium-walled
glass (WPI, 1BF100) and beveled on a Sutter Instruments beveler
(BV-10M).  Micropipettes were filled with 1.2 to 2M potassium acetate
and had resistances of 60-100 M$\Omega$ after beveling.  

All research procedures concerning the experimental animals and their
care adhered to the American Physiological Society's Guiding
Principles in the Care and Use of Animals, to the European Council
Directive 86/609/EEC and to European Treaties series no.\ 123, and
was also approved by the local ethics committee ``Ile-de-France Sud''
(certificate no.\ 05-003).

The dynamic-clamp technique (Robinson et al., 1993; Sharp et al.,
1993) was used to inject computer-generated conductances in real
neurons.  Dynamic-clamp experiments were run using the hybrid
RT-NEURON environment (Sadoc et al., 2009), which is a DSP-based
system using a modified version of the NEURON simulation environment
(Hines and Carnevale, 1997).  The dynamic-clamp protocol was used to
insert the fluctuating conductances underlying synaptic noise in
cortical neurons using the point-conductance model, similar to a
previous study (Destexhe et al., 2001).  According to
Eq.~(\ref{PCmodel}) above, the injected current is determined from
the fluctuating conductances $g_e(t)$ and $g_i(t)$ as well as from
the difference of the membrane voltage from the respective reversal
potentials, $I_{DynClamp} = -g_e \big( V - E_e \big) -g_i \big( V -
E_i \big)$.  The contamination of the measured membrane voltage $V$
by electrode artifacts due to simultaneous current injection was
avoided through the use of Active Electrode Compensation (AEC), a
novel, high resolution digital on-line compensation technique (Brette
et al. 2008).



\section*{Results}

We present here a method for extracting conductances by associating
the V$_m$ to a stochastic process, and which uses maximum likelihood
criteria to perform this extraction from single V$_m$ traces.  We
first describe the method, then test it successively using
computational models and real neurons {\it in vitro}.

\subsection*{The VmT method: extracting conductances from single V$_m$
traces} \label{vmt-sec}

The starting point of the method is to search for the ``most likely''
conductances parameters ( $g_{e0}$, $g_{i0}$, $\sigma_e$ and
$\sigma_i$) that are compatible with an experimentally observed V$_m$
trace.  We start from the point-conductance model
(Eqs.~\ref{PCmodel}--\ref{flucti}), which is discretized in time with
a step-size $\Delta t$.  Eq.~\ref{PCmodel} can then be solved for
$g_i^k$, which gives:
\begin{equation}
    \label{eq:gi_est}
    g_i^k = -\frac{C}{V^k-E_i} \left\{ \frac{V^k-E_L}{\tau_L} + 
      \frac{g_e^k (V^k - E_e)}{C} + \frac{V^{k+1} - V^k}{\Delta t} -
      \frac{I_{ext}}{C} \right\} ,
\end{equation}
where $\tau_L = C / G_L$.

Since the series $V^k$ for the voltage trace is known, $g_i^k$ has
become a function of $g_e^k$. In the same way, we solve
Eqs.~\ref{flucte}--\ref{flucti} for $\xi_s^k$, which have become 
Gaussian distributed random numbers,
\begin{equation}
    \label{eq:xi_est}
    \xi_s^k = \frac{1}{\sigma_s} \sqrt{\frac{\tau_s}{2 \Delta t}}
      \left( g_s^{k+1} - g_s^k \Big( 1 - \frac{\Delta t}{\tau_s} \Big)
      - \frac{\Delta t}{\tau_s} g_{s 0} \right),
\end{equation}
where $s$ stands for $e,i$.

There is a continuum of combinations $\{ g_e^{k+1}, g_i^{k+1}\}$ that
can advance the membrane potential from $V^{k+1}$ to $V^{k+2}$, each
pair occurring with a probability 
\begin{eqnarray}
    p^k &:=& p(g_e^{k+1}, g_i^{k+1} | g_e^k, g_i^k) = 
      \frac{1}{2 \pi} e^{-\frac{1}{2} (\xi_e^{k 2} + \xi_i^{k 2})} =
      \frac{1}{2 \pi} e^{-\frac{1}{4 \Delta t} X^k}, \\
    X^k &=& \frac{\tau_e}{\sigma_e^2} \left( g_e^{k+1} - 
      g_e^k \Big( 1 - \frac{\Delta t}{\tau_e} \Big) - 
      \frac{\Delta t}{\tau_e} g_{e 0} \right)^2 \\ 
    && + \frac{\tau_i}{\sigma_i^2} \left( g_i^{k+1} - 
      g_i^k \Big( 1 - \frac{\Delta t}{\tau_i} \Big) - 
      \frac{\Delta t}{\tau_i} g_{i 0} \right)^2. \nonumber
\end{eqnarray}
These expressions are identical to those derived previously for
calculating spike-triggered averages (Pospischil et al., 2007),
except that no implicit average is assumed here.

Thus, to go one step further in time, a continuum of pairs
$(g_e^{k+1}, g_i^{k+1})$ is possible in order to reach the (known)
voltage $V^{k+2}$. The quantity $p^k$ assigns to all such pairs a
probability of occurrence, depending on the previous pair, and the
voltage history. Ultimately, it is the probability of occurrence of
the appropriate random numbers $\xi_e^k$ and $\xi_i^k$ that relate
the respective conductances at subsequent time steps. It is then
straightforward to write down the probability $p$ for certain
conductance series to occur, that reproduce the voltage time course. 
This is just the probability for successive conductance steps to
occur, namely the product of the probabilities $p^k$:
\begin{equation}
    \label{eq:prob}
    p = \prod_{k = 0}^{n-1} p^k,
\end{equation}
given initial conductances $g_e^0$, $g_i^0$. However, again, there
is a continuum of conductance series
$$
 \{ g_e^l, g_i^l \}_{l = 1, \ldots, n+1},
$$ 
that are all compatible with the observed voltage
trace. We define a likelihood function $f(V^k, \theta)$, $\theta =
(g_{e0}, g_{i0}, \sigma_e, \sigma_i)$,  that takes into account all
of them with appropriate weight. We thus integrate Eq.~\ref{eq:prob}
over the unconstrained conductances $g_e^k$ and normalize by the
volume of configuration space:
\begin{equation}
    \label{eq:f_likelihood}
    f(V^k, \theta) = \frac{\int \prod_{k = 0}^{n-1} dg_e^k \; p}{\int \prod_{k = 0}^{n-1} dg_e^k \, dg_i^k \; p},
\end{equation}
where only in the nominator $g_i^k$ has been replaced by
Eq.~\ref{eq:gi_est}. This expression reflects the likelihood that a
specific voltage series $\{V^k\}$ occurs, normalized by the
probability, that {\it any} trace occurs. The most likely parameters
$\theta$ giving rise to $\{V^k\}$ are obtained by maximizing (or
minimizing the negative of) $f(V^k, \theta)$ using standard
optimization schemes (Press et al., 2007).

\subsection*{Test of the VmT method using model data}

\begin{figure}[h] 
\centerline{\psfig{figure=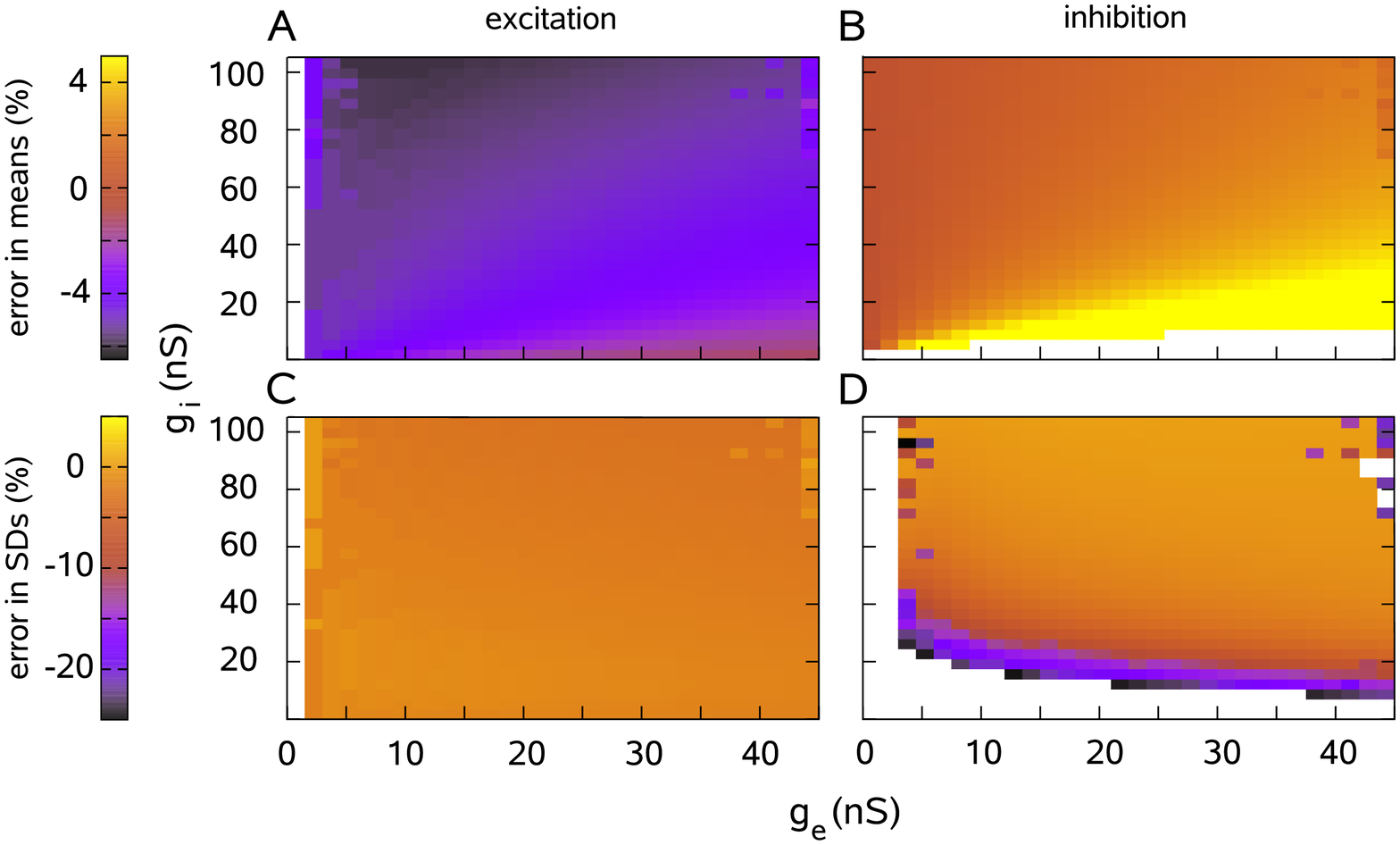,width=160mm}}

\caption{\label{fig:scan} Test of the single-trace VmT method using a
leaky integrator model.}

Each panel presents a scan in the ($g_{e 0}, g_{i 0}$)--plane. Color
codes the relative deviation between model parameters and their
estimates using the method (note the different scales for means/SDs).
The white areas indicate regions where the mismatch was larger,
$>$5\% for the means and $<$-25\% for SDs.  A. Deviation in the mean
of excitatory conductance ($g_{e 0}$).  B.  Same as A, but for
inhibition. C. Deviation in the SD of excitatory conductance.  D. 
Same as C, but for inhibition. In general the method works well,
except for a small band for the inhibitory SD.

\end{figure}

We tested the method in detail in its applicability to voltage traces
that were created using the same model (leaky integrator model). To
this end, we performed simulations scanning the ($g_{e 0}, g_{i
0}$)--plane, and subsequently tried to re-estimate the conductance
parameters used, but only from the V$_m$ activity.  For each
parameter set ($g_{e 0}, g_{i 0}$, $\sigma_e$, $\sigma_i$) the method
was applied to ten samples of 5000 data points (corresponding to
250~ms each) and the average was taken subsequently. The conductance
standard deviations (SDs) were chosen to be one third of the
respective mean values, other parameters were assumed to be known
during re-estimation ($C = 0.4$ nF, $g_L = 13.44$ nS, $E_L = -80$ mV,
$E_e$ = 0~mV, $E_i$ = -75~mV, $\tau_e = 2.728$ ms, $\tau_i = 10.49$
ms), the time step was $dt = 0.05$ ms. Also, we assumed that the
total conductance $g_{tot}$ was known -- this is the inverse of the
apparent input resistance, which is generally known.  This assumption
is not mandatory, but the estimation becomes more stable.  The
likelihood function given by Eq.~\ref{eq:f_likelihood} was thus only
maximized with respect to $g_{e0}, \sigma_e$ and $\sigma_i$. 
Fig.~\ref{fig:scan} summarizes the results obtained.

The mean conductances are well reproduced over the entire scan
region. An exception is the estimation of $g_{i 0}$ in the case where
the mean excitation exceeds inhibition by several-fold, a situation
which is rarely found in real neurons. The situation for the SDs is
different. While the excitatory SD is reproduced very well in the
whole area under consideration, this is not necessarily the case for
inhibition.  Here, the estimation is good for most parts of the
scanned region, but shows a considerable deviation along the left and
lower boundaries. These are regions where the transmembrane current
due to inhibition is weak, either because the inhibitory conductance
is weak (lower boundary) or because it is strong and excitation is
weak (left boundary), such that the mean voltage is close to the
inhibitory reversal potential and the driving force is small. In
these conditions, it seems that the effect of inhibition on the
membrane voltage cannot be distinguished from that of the leak
conductance.  

\begin{figure}[h] 
\centerline{\psfig{figure=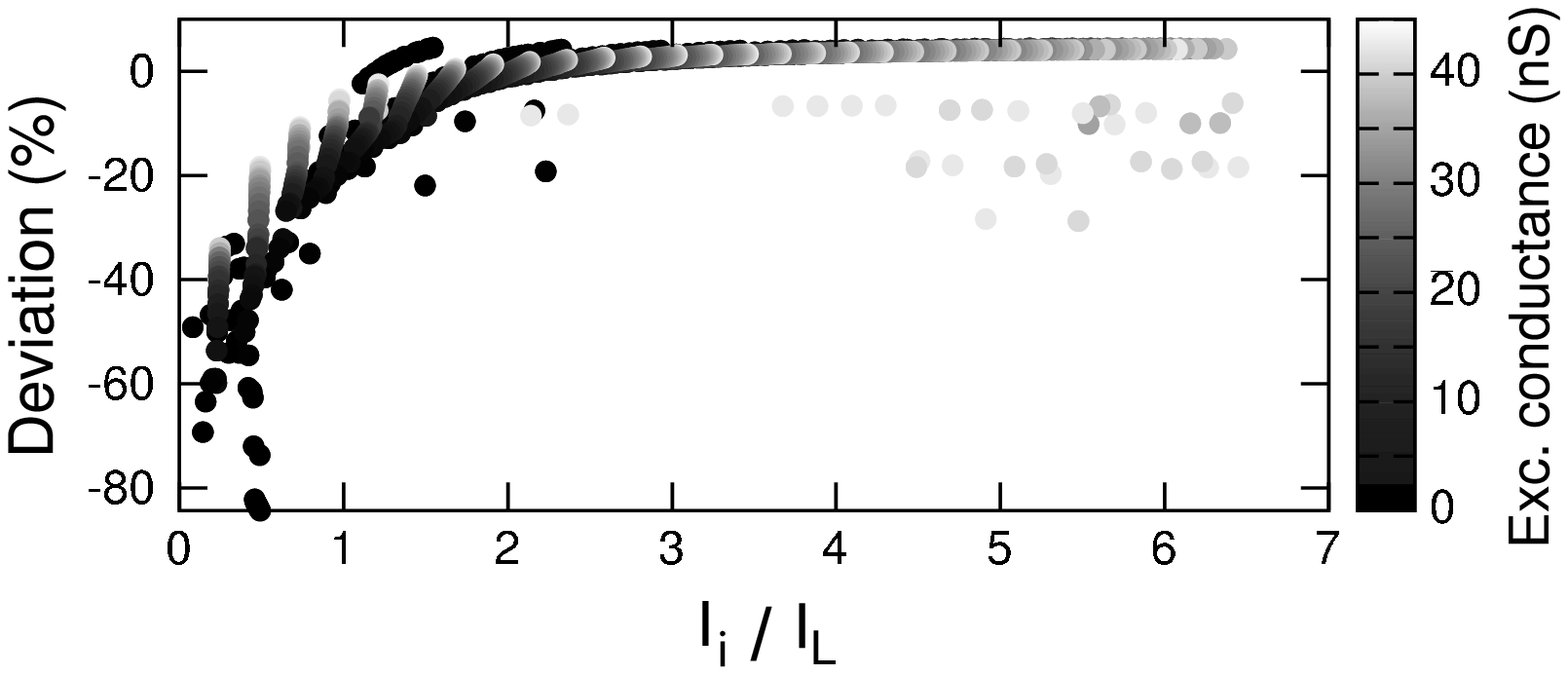,width=110mm}}

\caption{\label{fig:inh-leak} The estimation error depends on the
ratio of inhibitory and leak conductances.}

The relative deviation between the parameter $\sigma_i$ in the
simulations and its re-estimated value is shown as a function of the
ratio of the currents due to inhibitory and leak conductances.  The
estimation fails when the inhibitory component becomes too small. 
The same data as in Fig.~\ref{fig:scan} are plotted: different dots
correspond to different pairs of excitatory and inhibitory
conductances ($g_{e 0}$, $g_{i 0}$), and the dots are colored
according to the excitatory conductance (see scale).  Details about
the duration of injections etc. are given in the text.

\end{figure}

This point is illustrated in Fig.~\ref{fig:inh-leak}.  The relative
deviation between $\sigma_i$ in the model and its re-estimation
depends on the ratio of the transmembrane current due to inhibitory
($I_i$) and leak ($I_L$) conductance.  The estimation fails when the
inhibitory current is smaller or comparable to the leak current, but
it becomes very reliable as soon as the ratio $I_i/I_L$ becomes
larger than 1.5--2.  Some points, however, have large errors although
with dominant inhibition.  These points have strong excitatory
conductances (see gray scale) and correspond to the upper right
corner of Fig.~\ref{fig:scan}D.  The error is due to aberrant
estimates for which the predicted variance is zero; in principle such
estimates could be detected and discarded, but no such detection was
attempted here.  Besides these particular combinations, the majority
of parameters with strong inhibitory conductances gave acceptable
errors.  We conclude that the estimation of the conductance variances
will be most accurate in high-conductance states, where inhibitory
conductances are strong and larger than the leak conductance.

\subsection*{Effect of recording noise}

\begin{figure}[h] 
\centerline{\psfig{figure=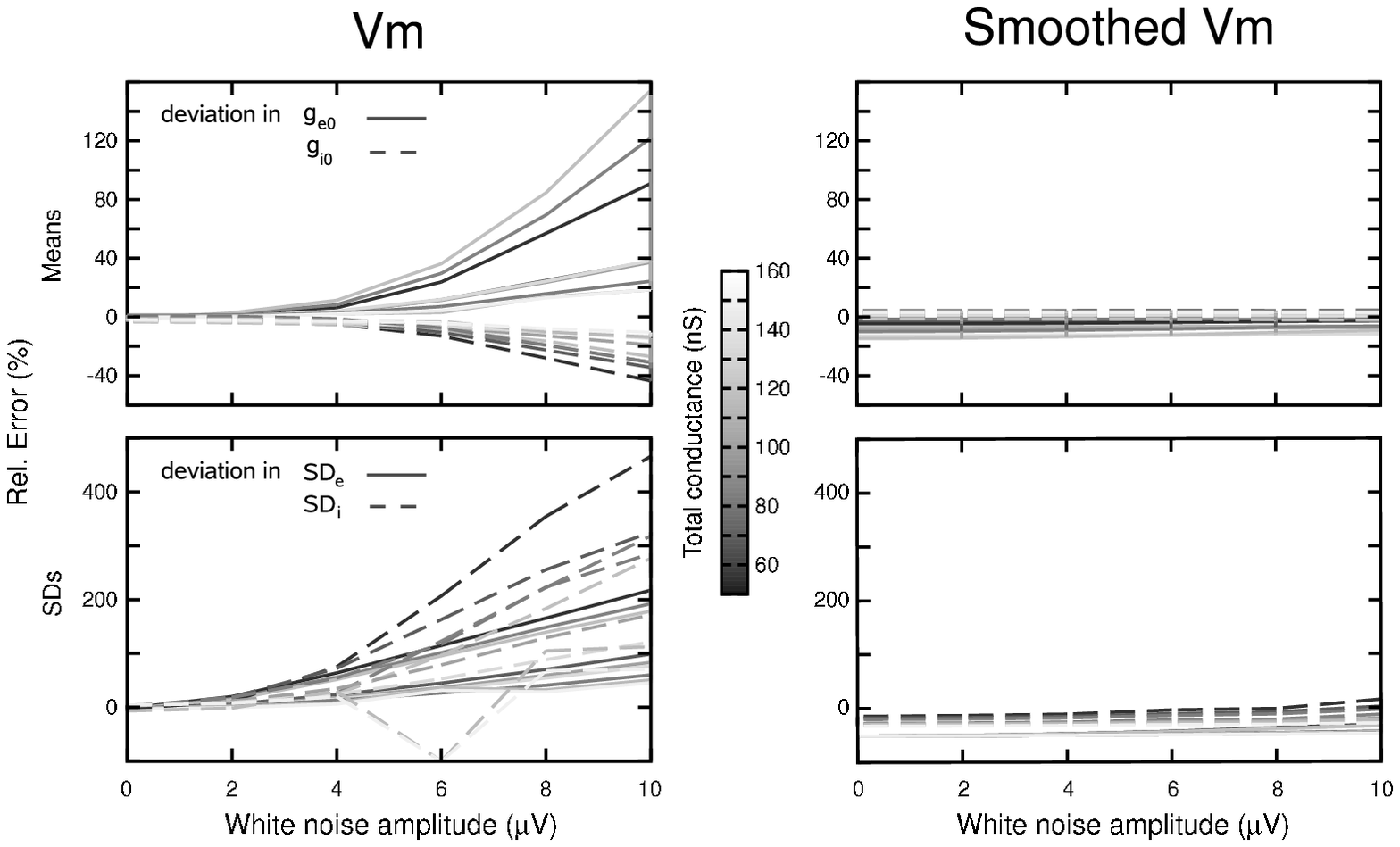,width=160mm}}

\caption{\label{fig:white-noise} Error of the VmT estimates following
addition of white noise to the voltage trace.}

Gaussian white noise was added to the voltage trace of the
model, and the VmT method was applied to the V$_m$ trace obtained
with noise, to yield estimates of conductances and variances.  Left:
relative error obtained in the estimation of $g_{e 0}$ and $g_{i 0}$
(upper panel), as well as $\sigma_e$ and $\sigma_i$ (lower panels). 
Right: same estimation, but the V$_m$ was smoothed prior to the VmT
estimate (Gaussian filter with SD of 1 data point).  In both cases,
the relative error is shown as a function of the white noise
amplitude.  Different curves correspond to different pairs ($g_{e
0}$, $g_{i 0}$).  The errors on the estimates for both mean
conductance and SD increase with the noise.  The coloring of the
curves as a function of the total conductance (see scale) shows that
the largest errors generally occur for the low-conductance regimes. 
The error was greatly diminished by smoothing (right panels).

\end{figure}

The unavoidable presence of recording noise may present a problem in
the application of the method to recordings from real neurons. 
Fig.~\ref{fig:white-noise} (left) shows how low-amplitude white noise
added to the voltage trace of a leaky integrator model impairs the
reliability of the method.  A Gaussian-distributed white noise was
added to the voltage trace at every time step, scaled by the
amplitude given in the abscissa.  Different curves correspond to
different pairs ($g_{e 0}$, $g_{i 0}$) colored as a function of the
total conductance. The noise has an opposite effect on the estimation
of the conductance mean values.  While the estimate of excitation
exceeds the real parameter value, for inhibition the situation is
inverted.  However, one has to keep in mind that both parameters are
not estimated independently, but their sum is kept fixed. In
contrast, the estimates for the conductance SDs always exceed the
real values, and they can deviate by almost 500\% for a noise
amplitude of 10 $\mu$V.  Here, the largest errors generally
correspond to lowest conductance states. Clearly, in order to apply
the method to recordings from real neurons, one needs to restrain
this noise sensitivity.  

Fortunately, this noise sensitivity can be reduced by standard noise
reduction techniques.  For example, preprocessing and smoothing the
data using a Gaussian filter greatly diminishes the amplitude of the
noise, and consequently improves the estimates according to the new
noise amplitude (see Fig.~\ref{fig:white-noise}, right panels).  Too
much smoothing, however, may result in altering the signal itself,
and may introduce errors.  It is therefore preferable to use
smoothing at very short time scales (SD of 1--4 data points,
depending on the sampling rate). In the next sections, we
preprocessed the experimental voltage traces with a Gaussian filter
with a SD of 3 data points.

\subsection*{Application of the VmT method to {\it in vitro} data}

\begin{figure} 
\centerline{\psfig{figure=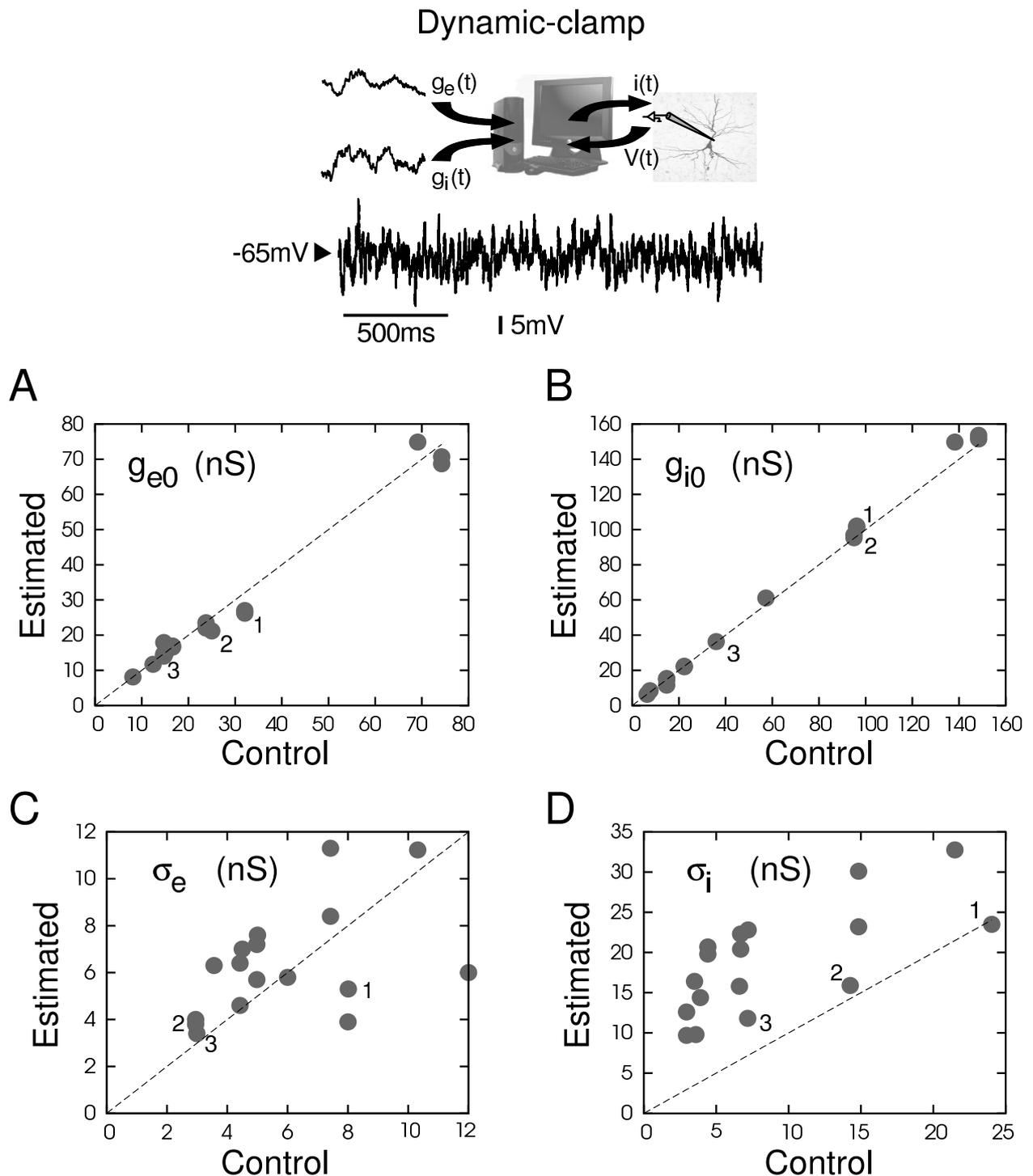,width=165mm}}

\caption{\label{vmt} Dynamic-clamp test of the VmT method to extract
conductances from guinea-pig visual cortical neurons {\it in vitro}.}

Fluctuating conductances of known parameters were injected in
different neurons using dynamic-clamp, and the V$_m$ activity
produced was analyzed using the single-trace VmT method.  Each plot
represents the different conductance parameters extracted from the
V$_m$ activity: $g_{e0}$ (A), $g_{i0}$ (B), $\sigma_e$ (C) and
$\sigma_i$ (D).  The extracted parameter (Estimated) is compared to
the value used in the conductance injection (Control). While in
general the mean conductances are matched very well, the estimated
SDs show a large spread around the target values.  Nevertheless,
during states dominated by inhibition (see indexed symbols), the
estimation was acceptable.

\end{figure}

We next tested the method on {\it in vitro} recordings (n=5 cortical
neurons, 17 injections) using dynamic clamp experiments
(Fig.~\ref{vmt}).  As in the model, the stimulus consisted of two
channels of fluctuating conductances representing excitation and
inhibition.  The conductance injection spanned values from
low-conductance (of the order of 5-10~nS) to high-conductance states
(50-160~nS).  It is apparent from Fig.~\ref{vmt} that the mean values
of the conductances ($g_{e0}$, $g_{i0}$) are well estimated, as
expected because the total conductance is known in this case.

\begin{figure} 
\centerline{\psfig{figure=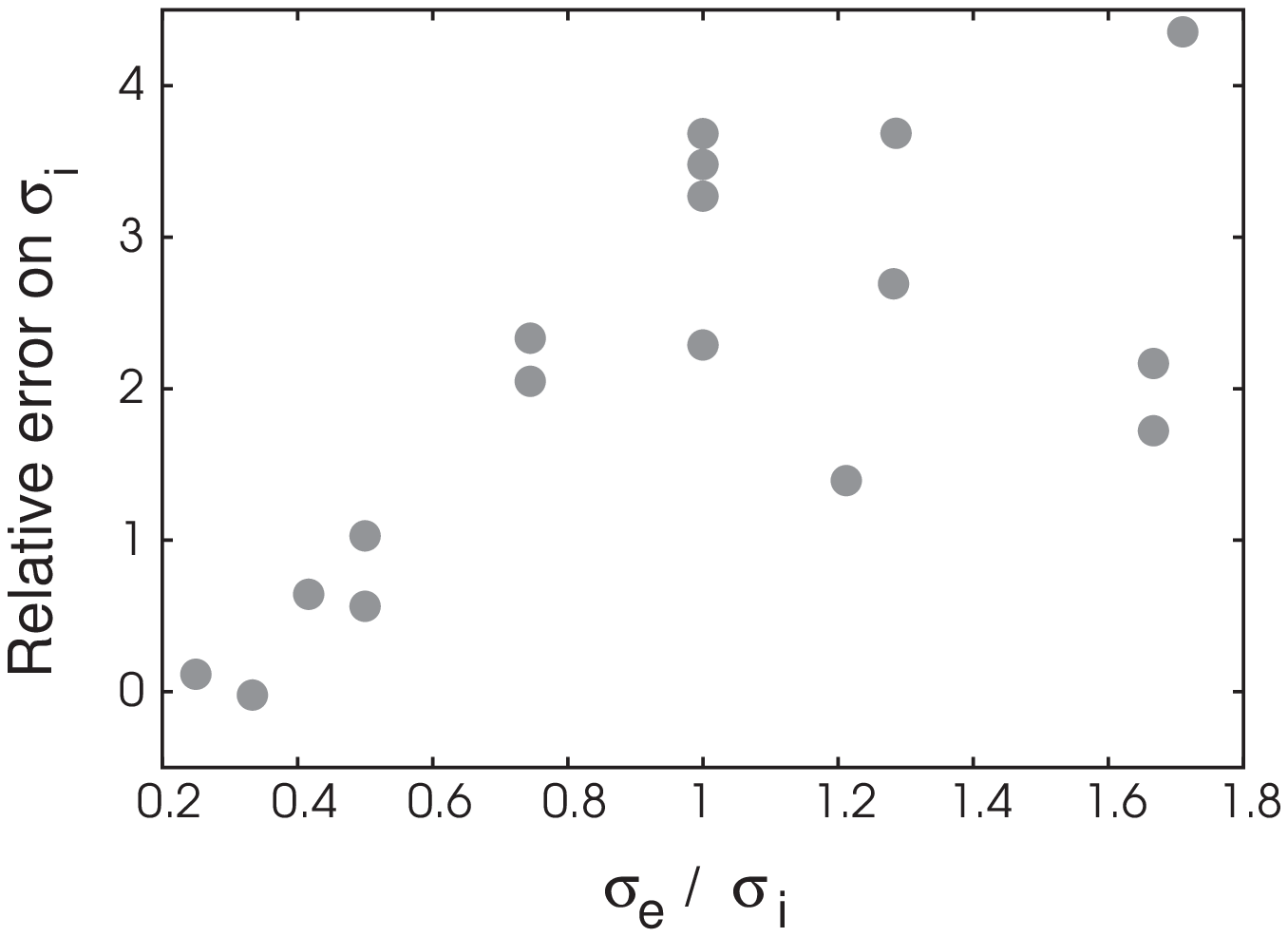,width=120mm}}

\caption{\label{relerror} Relative error on inhibitory variance is
high only when excitatory fluctuations dominate.}

The relative mean-square error on $\sigma_i$ is represented as a
function of the $\sigma_e / \sigma_i$ ratio.  The error is
approximately proportional to the ratio of variances.  The same data
as in Fig.~\ref{vmt} were used.

\end{figure}

However, the estimation is subject to larger errors for the standard
deviations of the conductances ($\sigma_e$, $\sigma_i$).  In
addition, the error on estimating variances is also linked to the
accuracy of the estimates of synaptic time constants $\tau_e$ and
$\tau_i$, similar to the VmD method (see discussion in Piwkowska et
al., 2008). Interestingly, for some cases, the estimation works quite
well (see indexed symbols in Fig.~\ref{vmt}).  In the pool of
injections, there are three cases that represent a cell in a
high-conductance state, i.e. the mean inhibitory conductances are
roughly three times greater than the excitatory ones, and the
standard deviations obey a similar ratio.  For these trials, the
estimate comes close to the values used during the experiment. 
Indeed, we found that the relative error on $\sigma_i$ is roughly
proportional to the ratio $\sigma_e / \sigma_i$ for ratios smaller
than 1, and tends to saturate for larger ratios
(Fig.~\ref{relerror}).  In other words, the estimation has the lowest
errors when inhibitory fluctuations dominate excitatory fluctuations.
A recent estimate of conductance variances in cortical neurons of
awake cats reported that $\sigma_i$ is larger than $\sigma_e$ for the
vast majority of cells analyzed (Rudolph et al., 2007).  The same was
also true for anesthetized states (Rudolph et al., 2005), suggesting
that the VmT method should give acceptable errors in practical
situations {\it in vivo}.



\section*{Discussion}

In this paper, we have introduced and tested a new method to estimate
conductance mean and variance from single V$_m$ traces.  Traditional
methods to estimate conductances require to perform current-clamp or
voltage-clamp recordings at different voltages (Rudolph et al., 2004;
Monier et al., 2008).  If the membrane has intrinsic
voltage-dependent conductances, the use of different voltage levels
will necessarily cause important errors.  This is particularly
critical {\it in vivo}, where pharmacological manipulations to
suppress currents are usually very limited.

Several methods exists that can be applied to single V$_m$ traces;
for example, one can calculate power spectral densities (PSD) of the
V$_m$.  The fitting of analytical expressions of the V$_m$ PSD would
allow in principle to estimate the conductance parameters, since they
appear in these expressions.  However, this analysis would require to
know the effective membrane time constant, and therefore the total
conductance.  In such a case, since the average V$_m$ is known, the
conductances can be estimated by simply solving the membrane equation
at steady-state, so the PSD presents no useful means of estimating
conductances.  Nevertheless, fitting theoretical expressions to
experimental PSDs can in principle yield estimates of other
parameters, such as the decay time constants of synaptic
conductances, although such estimates can be subject to important
errors (see Piwkowska et al., 2008 for an assessment of this method).

A second type of method applicable to single V$_m$ traces is to
estimate the spike-triggered average (STA) of synaptic conductances
(Pospischil et al., 2007).  The principle of this method is to search
for the most probable conductance traces that account for the V$_m$
STA.  These traces are calculated by maximum likelihood following a
discretization of the time axis.  The performance of this method was
tested using computational models and dynamic-clamp experiments
(Pospischil et al., 2007; Piwkowska et al., 2008).  It was found to
be excellent, and the STA method was applied to estimate the optimal
conductance patterns preceding spikes {\it in vivo} (Rudolph et al.,
2007).  However, this method requires a prior knowledge of the
conductance parameters ($g_{e0}$, $g_{i0}$, $\sigma_e$, $\sigma_i$). 
In practice, it is thus necessary to apply the VmD method beforehand,
which in turn requires recording at several V$_m$ levels.

We proposed here a method that merges the above concepts.  The VmT
method can estimate the same conductance parameters as the VmD
method ($g_{e0}$, $g_{i0}$, $\sigma_e$, $\sigma_i$), but using a
maximum likelihood approach similar to the STA estimates.  The
analysis can thus be realized from single V$_m$ levels, thereby
limiting the bias caused by voltage-dependent currents.  The VmT
method does not rely on static properties of the membrane voltage,
like mean and standard deviation, in order to determine the shape
of the conductance distributions. Instead, it exploits the
dynamical information hidden in the V$_m$ time course.  Not only
the step sizes $\Delta t$ are important, but also at which voltage
level they occur. The likelihood function is highly sensitive to
the conductance fluctuations and, constraining the total
conductance, also to the conductance mean values.  

A word of caution is needed here about the biological significance of
the conductances estimated by the VmT method. Like any other method
to extract conductances from V$_m$ activity, the conductance
estimated here refers to the conductance visible at the site of the
V$_m$ recording, in general the soma.  This somatic conductance can
be very different -- and is in general smaller -- than the
conductance present at the site of the synaptic receptors, mostly the
dendrites.  This is especially critical for neurons with extended
dendritic morphologies, such as pyramidal neurons, which experience
considerable dendritic attenuation.  However, the visible somatic
conductances are close to the conductances participating to the
interactions involved in generating the action potential, because the
region for spike initiation (presumably the initial part of the axon;
see review by Stuart et al., 1997) is electrotonically close to the
soma.  To yield estimates closer to the conductances present in
dendrites, recordings must be done using drugs such as QX-314 and
cesium, to reduce the leak conductance and thereby lead to more
compact neurons, with more limited dendritic attenuation.  The
conductance contribution due to spikes (somatic or dendritic) should
be avoided by using exclusively subthreshold activity.

Similarly, like all other methods for the analysis of synaptic
conductances, the VmT method rests on the assumption that the range
of V$_m$ analyzed is linear, and does not contain voltage-dependent
conductances.  Nevertheless, it is possible to follow a similar
maximum-likelihood procedure by explicitly including fast
voltage-dependent conductances in the model.  The VmT method also
requires knowledge of the passive parameters of the recorded neuron
($G_L$, $C$...) which may be difficult to obtain {\it in vivo} (see
Piwkowska et al., 2008 for a discussion of this point).

A second word of caution concerns the significance of conductance
variances.  The conductance variances estimated using the VmT method
are specific to the ``point-conductance'' model, in which the V$_m$
activity is assumed to result from the action of two stochastic
synaptic conductances (Destexhe et al., 2001).  In particular, one of
the assumptions was that the excitatory and inhibitory conductances
are Gaussian-distributed.  Estimates of variances of course depend on
this assumption.  Including asymmetric conductance distributions is
possible but would substantially complicate the model.  The
mathematical simplicity of this model has made it possible to design
a series of methods to analyze the V$_m$ activity (reviewed in
Piwkowska et al., 2008).  The VmT method is a natural extension of
this approach.

The tests of the VmT method on model data were encouraging, but also
pointed to weaknesses of the method in a regime where the
transmembrane current due to inhibitory conductances is small
compared to the leak current.  In this case, the inhibitory
fluctuations are not reliably resolved.  Furthermore, the presence of
recording noise can disturb the estimation significantly.  Though
smoothing the voltage trace could in some cases re-establish a good
result (Fig.~\ref{fig:white-noise}), more sophisticated procedures
should be considered to eliminate this sensitivity to noise.  For
example, one could introduce to Eq.~\ref{PCmodel} an additive noise
term, whose SD would become an additional parameter that has to be
estimated.  Thus, in principle, if this weak inhibition regime can be
avoided, the VmT method should be applicable to real neurons.  In any
case, to be on the safe side, it is preferable to use the VmT method
on V$_m$ traces with the lowest possible level of instrumental noise.

We also tested the VmT method by using dynamic-clamp injection of
stochastic conductances.  In this case, controlled conductances are
injected while the V$_m$ activity is being recorded, allowing to
compare the VmT estimates from the V$_m$, to the conductances
actually injected.  These tests confirmed the weakness of the method
in the low-inhibition regime, but also revealed a good performance of
the VmT method during HC states. Since in cortical neurons {\it in
vivo} the membrane potential is usually well above the inhibitory
reversal potential, and since inhibitory conductances tend to
dominate over excitatory ones (Borg-Graham et al., 1998; Rudolph et
al., 2007; note that dendritic recordings may be dominated by
excitatory inputs), we anticipate that this method should provide
good conductance estimates from {\it in vivo} recordings.


\subsection*{Acknowledgments}

This research was supported by the Centre National de la Recherche
Scientifique (CNRS, France), the Agence Nationale de la Recherche
(ANR, France), Future and Emerging Technologies (FET, European Union)
and the Human Frontier Science Program (HFSP).  Additional
information is available at \url{http://cns.iaf.cnrs-gif.fr}

\clearpage
\section*{References}





\begin{enumerate}

\small

\item Baranyi A, Szente MB, Woody CD. (1993) Electrophysiological
characterization of different types of neurons recorded in vivo in
the motor cortex of the cat. II. Membrane parameters, action
potentials, current-induced voltage responses and electrotonic
structures.  J.  Neurophysiol. 69: 1865-1879.

\item Borg-Graham, L.J., Monier, C., Fr\'egnac, Y. (1998) Visual
input evokes transient and strong shunting inhibition in visual
cortical neurons. Nature 393: 369-373.

\item Braitenberg, V., Sch\"uz, A. (1998)  {\it Cortex: statistics
and geometry of neuronal connectivity} (2nd edition),
Springer-Verlag, Berlin.

\item Brette R, Piwkowska Z, Monier C, Rudolph-Lilith M, Fournier J,
Levy M, Fr\'egnac Y, Bal T, Destexhe A  (2008) High-resolution
intracellular recordings using a real-time computational model of the
electrode. Neuron 59: 379-391.

\item DeFelipe, J., Fari\~nas, I. (1992)  The pyramidal neuron of the
cerebral cortex: morphological and chemical characteristics of the
synaptic inputs. Prog. Neurobiol. 39: 563-607.

\item Destexhe, A. (2007) High-conductance state.  Scholarpedia
2(11): 1341 \\
\url{http://www.scholarpedia.org/article/High-Conductance\_State}

\item Destexhe A, Rudolph M, Fellous J-M, Sejnowski TJ. (2001)
Fluctuating synaptic conductances recreate in-vivo--like activity in
neocortical neurons.  Neuroscience 107: 13-24.

\item Destexhe A, Rudolph, M, Par\'e D (2003) The high-conductance
state of neocortical neurons in vivo.  Nature Reviews Neurosci. 4:
739-751.

\item Evarts, E.V. (1964) Temporal patterns of discharge of pyramidal
tract neurons during sleep and waking in the monkey. J. 
Neurophysiol. 27: 152-171.

\item Gillespie, D.T. (1996) The mathematics of Brownian motion and
Johnson noise. Am. J. Phys. 64: 225-240.

\bibitem{Hines97} Hines, M.L. and N.T. Carnevale. (1997)  The NEURON
simulation environment. Neural Computation 9: 1179-1209.

\item Longtin, A. (2008) Neuronal noise.  Scholarpedia (in press) \\
\url{http://www.scholarpedia.org/article/Neuronal\_noise}

\item Matsumara M, Cope T, Fetz EE. (1988) Sustained excitatory
synaptic input to motor cortex neurons in awake animals revealed by
intracellular recording of membrane potentials.  Exp.  Brain Res. 
70: 463-469.

\item Monier C, Fournier J, Fr\'egnac Y. (2008) In vitro and in
vivo measures of evoked excitatory and inhibitory conductance
dynamics in sensory cortices.  J. Neurosci. Meth. (in press)

\item Piwkowska, Z., Pospischil, M., Brette, R., Sliwa, J.,
Rudolph-Lilith, M., Bal, T., Destexhe, A. (2008)  Characterizing
synaptic conductance fluctuations in cortical neurons and their
influence on spike generation. J. Neurosci. Methods 169: 302-322.

\item Pospischil M, Piwkowska Z, Rudolph M, Bal T, Destexhe A. (2007)
Calculating event-triggered average synaptic conductances from the
membrane potential.  J. Neurophysiol. 97: 2544-2552.

\item Press W.H., Flannery B.P., Teukolsky S.A., Vetterling W.T. 
(2007) {\it Numerical Recipes: The Art of Scientific Computing} (3rd
edition). Cambridge University Press, Cambridge UK.

\item Robinson H.P. and Kawai N. (1993) Injection of digitally
synthesized synaptic conductance transients to measure the
integrative properties of neurons. J. Neurosci. Methods 49: 157-165.

\item Rudolph M, Destexhe A. (2003c) Characterization of subthreshold
voltage fluctuations in neuronal membranes. Neural Computation 15:
2577-2618.

\item Rudolph M, Destexhe A. (2005) An extended analytic expression
for the membrane potential distribution of conductance-based synaptic
noise.  Neural Computation 17: 2301-2315.

\item Rudolph M, Piwkowska Z, Badoual M, Bal T, Destexhe A. (2004) A
method to estimate synaptic conductances from membrane potential
fluctuations. J. Neurophysiol. 91: 2884-2896.

\item Rudolph, M., Pelletier, J-G., Par\'e, D. and Destexhe, A. 
(2005) Characterization of synaptic conductances and integrative
properties during electrically-induced EEG-activated states in
neocortical neurons in vivo. J. Neurophysiol. 94: 2805-2821.

\item Rudolph M, Pospischil M, Timofeev I, Destexhe A. (2007)
Inhibition controls action potential generation in awake and sleeping
cat cortex.  J Neurosci. 27: 5280-5290.  

\item Sadoc G, Le Masson G, Foutry B, Le Franc Y, Piwkowska Z,
Destexhe A and Bal T. (2009) Recreating {\it in vivo}--like activity
and investigating the signal transfer capabilities of neurons:
Dynamic-clamp applications using real-time NEURON.  In: {\it
Dynamic-clamp: From Principles to Applications}, Edited by Destexhe A
and Bal T, Springer, new York, in press.

\bibitem{Sharp93} Sharp, A.A., O'Neil, M.B., Abbott, L.F. and
Marder, E. (1993) The dynamic clamp: artificial conductances in
biological neurons. Trends Neurosci. 16: 389-394.

\item Steriade, M., McCarley, R.W. (1990)  {\it Brainstem Control of
Wakefulness and Sleep}, Plenum Press, New York.

\item Steriade M, Timofeev I, Grenier F. (2001) Natural waking and
sleep states: a view from inside neocortical neurons.  J.
Neurophysiol. 85: 1969-1985.

\item Stuart, G., Spruston, N., Sakmann, B. and Hausser, M.  (1997)
Action potential initiation and backpropagation in neurons of the
mammalian CNS. Trends Neurosci. 20: 125-131.

\end{enumerate}

\clearpage
\section*{Figures}

\end{document}